# A Carbon Corrosion Model to Evaluate the Effect of Steady State and Transient Operation of a Polymer Electrolyte Membrane Fuel Cell

Arun Pandy[1], Zhiwei Yang[2], Mallika Gummalla[2*], Vadim V. Atrazhev[3,4],

Nikolay Yu. Kuzminyh [3,4], Vadim I. Sultanov [4], Sergei Burlatsky[2]

[1] UTC Power Corporation, 195 Governors Highway, South Windsor, CT 06074, USA

[2] United Technologies Research Center, 411 Silver Lane, East Hartford, CT 06108, USA

[3] Russian Academy of Science, Institute of Biochemical Physics, Kosygin str. 4, Moscow, 119334, Russia

[4] Science for Technology LLC, Leninskiy pr-t 95, Moscow, 119313, Russia

**Abstract**

A carbon corrosion model is developed based on the formation of surface oxides on carbon and platinum of the polymer electrolyte membrane fuel cell electrode. The model predicts the rate of carbon corrosion under potential hold and potential cycling conditions. The model includes the interaction of carbon surface oxides with transient species like OH radicals to explain observed carbon corrosion trends under normal PEM fuel cell operating conditions. The model prediction agrees qualitatively with the experimental data supporting the hypothesis that the interplay of surface oxide formation on carbon and platinum is the primary driver of carbon corrosion.

[*] Corresponding author

**Introduction**

Carbon is commonly used as material for catalyst supports, gas-diffusion media and bipolar plates in Proton Exchange Membrane (PEM) fuel cells, though it is well-known that carbon is thermodynamically unstable in PEM fuel cell cathode environments. The current PEM fuel cell cathodes typically operate at temperatures in the range of 60ºC – 85ºC and potentials in the range of 0.5 – 0.95 V (vs. RHE), which is significantly more anodic than the equilibrium potential for carbon oxidation to carbon dioxide (0.207V vs. RHE). However, the kinetics of carbon oxidation under PEM operational conditions is relatively slow. This slow oxidation kinetics makes carbon an appropriate material for PEM electrodes.

Carbon corrosion occurs at different rates under various fuel cell operating conditions. A few distinct conditions that can lead to extensive carbon corrosion and catastrophic performance decay in a short period of time have been identified. One such condition occurs during start/stop operations when air leaks into or is present in the anode gas channels and creates potential variation in the platform[1–3]. Another condition that leads to carbon corrosion is fuel starvation induced by flow mal-distribution in the individual cells of the stack[4]. In both cases, undesirable oxygen reduction reaction (ORR) at the anode decreases the potential of ionomer keeping high potential difference between carbon and ionomer (close to equilibrium potential of ORR). That results in high potential difference (~1.4 V) between carbon and ionomer at the opposite cathode and accelerated carbon corrosion at this potential[1-4]. The corresponding system strategies to mitigate have also been proposed and demonstrated[5].

Under normal PEM fuel cell operating conditions, carbon corrosion occurs at a less perceptible rate though accelerated by platinum[6–10] and water. Although this impact is less severe than for hydrogen starvation and startup, it is still large enough to cause the degradation of PEMFC cell performance after long-term operation[11]. The carbon corrosion mechanism and approaches of modeling under normal PEM fuel cell operating conditions have received increased attention due to long-term performance and durability requirements on the current fuel cell systems. Few possible routes for carbon corrosion reported in refs[9,12–14] suggest that the transient species such as ·OH and ·OOH radicals formed on the surface of the Pt catalyst spill over to the carbon, resulting in formation of carbon oxide groups, which then proceed to CO and $CO_2$.

Under normal PEM fuel cell operating conditions, carbon corrosion does not occur on anode side since the anode potential is always lower than the equilibrium potential for carbon oxidation (0.207V vs. RHE). Oxidation of carbon is believed to happen only on cathode side, especially in electrodes and electrode/gas diffusion layer (GDL) interface because of the presence of platinum catalyst. Oxidation of carbon into $CO_2$ is known to be strongly dependent on carbon morphology and functional groups on its surface. The morphology of carbon surface could vary from highly-stable graphite structure to less-stable amorphous structure. Carbon atoms on the surface could also covalently bind with other atoms to form a wide variety of species, including C-H, phenols, carbonyls, carboxylic acids, ethers and quinone etc.[10]. These species have different thermodynamic oxidation-reduction stability within the range of PEM fuel cell operating conditions generally referred to as potential, temperature, relative humidity (RH) and oxygen concentration. Therefore, multiple factors simultaneously impact carbon corrosion under normal

PEM fuel cell operating conditions making it very difficult to delineate the contributions from each of the operating condition. Hence, several controlled tests are needed to elucidate the fundamentals of carbon corrosion. While the carbon surface changes are hard to detect *in situ*, the rate of carbon loss is detectable by monitoring $CO_2$ emission, especially during transient operating conditions that cause potential changes on electrodes[15,16].

The focus of this paper is to present a model that captures the carbon loss and performance loss due to potential cycling or load cycling. Towards this objective, a series of experiments were conducted to elucidate the fundamentals of carbon corrosion and guide the model development. A model that captures carbon corrosion occurring under such transient conditions is described in the following sections.

**Experimental**

*Electrode Preparation and Evaluation:* Pt/carbon catalyst (TEC10E50E, 46.7% Pt on Ketjen black carbon, Tanaka Kikinzoku Kogyo KK, Japan) was dispersed in isopropyl alcohol/D.I. water (60:40 volume ratio) mixture. Nafion® dispersion (5 wt %, 1100 equivalent weight; Dupont Fluoroproducts, Wilmington, DE) was added to form catalyst-ionomer ink with the catalyst / Nafion® weight ratio of 79 / 21. The resultant ink was screen printed on Teflon® film (127 micron thickness) to form decals, which had Pt loading of 0.4±0.04 mg-Pt/cm$^2$. The ink-coated decals were dried in a $N_2$ protected chamber at room temperature and 30% relative humidity overnight. The dried decals were then hot-pressed at the same time onto both sides of Nafion®-111 membranes ($H^+$ form, as received) under 130°C and 450 Psi for 5 minutes. The resultant membrane-electrode assembly (MEA) was sandwiched between two pieces of carbon papers with a micro-porous layer (SGL-25BC, SGL Carbon Group, Germany) as gas diffusion media, and assembled into single-cell hardware (25cm$^2$ active area, Fuel Cell Technologies Inc., Albuquerque, NM). The assembled single cell was conditioned by scanning current between 0.1 A/cm$^2$ and 1.5 A/cm$^2$ (or 0.35V of cell voltage) in steps of 0.1A/cm$^2$ with 5 min hold at 75°C with 92% RH inlet H$_2$ and air (1 slpm and 2 slpm, respectively) for 48 hours.

In different cell configurations, pure nano-Pt particle was mixed with Nafion® dispersion to form the electrode ink with the Pt / Nafion® weight ratio of 92/8. The resultant ink was used to prepare the cathode electrode. Besides SGL-25BC, carbon paper without micro-porous layers (U107, SGL Carbon Group, Germany) was also used as cathode diffusion media in a different cell setup.

***Reactants***: Hydrogen (H$_2$) used in this study was generated on-site by a Hydrogen Generator (HOGEN® 40, Proton Energy Systems Inc.). Nitrogen (N$_2$) and air used for carbon corrosion tests were Volatile-Organic-Compound (VOC) Free grade from Praxair, Inc. (Part number #: NI 5.0VC and AI 0.0VC respectively).

***Online Carbon Corrosion Detection:*** The cathode exhaust of the fuel cell was introduced into a Nafion® based gas dryer (Perma Pure LLC, Toms River, NJ, USA) to reduce the moisture concentration to ~ 2000 ppm. Then the dried cathode gas stream was introduced into a photo-acoustic infrared gas analyzer (Bruel & Kjaer Multi-gas Monitor, Type 1302) to measure the CO$_2$ concentration. The detection limit of CO$_2$ concentrations in dried gas stream was sub-ppm level.

*Identification of CO$_2$ Source from Fuel Cell:* Various carbon-containing components in fuel cells could be oxidized into CO$_2$ under the fuel cell operating conditions. However, the presence of platinum catalysts on high surface area carbon is believed to be the dominant source for the carbon loss. The carbon corrosion of GDLs and carbon bipolar plates is believed to be largely negligible even over a prolonged operation. To verify this hypothesis three cells with the same anode configuration but different cathode configurations were tested: Configuration 1 had carbon in the cathode electrode (TEC10E50E) and micro-porous layer of the GDL (SGL-25BC), Configuration 2 had carbon in micro-porous layer only (nano Pt black catalyst electrode), and Configuration 3 had no carbon in the electrode or micro-porous layer (nano Pt black catalyst electrode + U107 GDL). The three configurations were tested at 75ºC, 95% RH, 0.15 slpm H$_2$ & 0.36 slpm N$_2$ on anode and cathode respectively. The net carbon dioxide measured from the three cells at various cyclic conditions is shown in Figure 1.

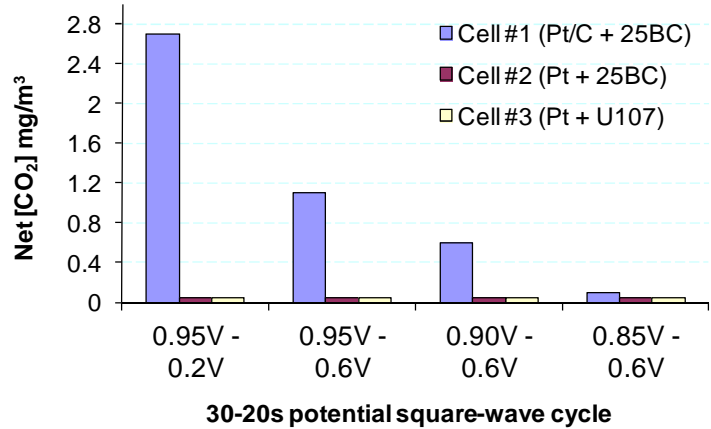

Figure 1. Increment of CO$_2$ concentration in cathode exhausts during the potential square-wave cycling (50s/cycle, 30s at upper voltages and 20s at low voltages)

While the steady state $CO_2$ emissions from the three cell configurations were identical at constant potential holds, the difference were evident when square wave cycled to different upper potentials. Square-wave potential cycling is believed to accelerate cell decay[17]. The $CO_2$ concentration in the cathode exhaust significantly increased for the configuration 1 (with carbon support in cathode electrode), but no statistically significant increase was observed for configurations 2 and 3. This result clearly shows that carbon corrosion mainly occurred on carbon support in the cathode electrode. The carbon corrosion rates in micro-porous layer of SGL-25BC as well as carbon paper (U107) were too small to detect in the current experiments. Configuration 1 is the baseline case used in this study.

**Carbon Corrosion model**

The carbon corrosion model that is developed here predicts the $CO_2$ released from the fuel cell, as a function of potential and potential cycling conditions assuming all the carbon loss was from the cathode electrode carbon. The model captures the effect of water fraction / relative humidity and upper / lower potential limits.

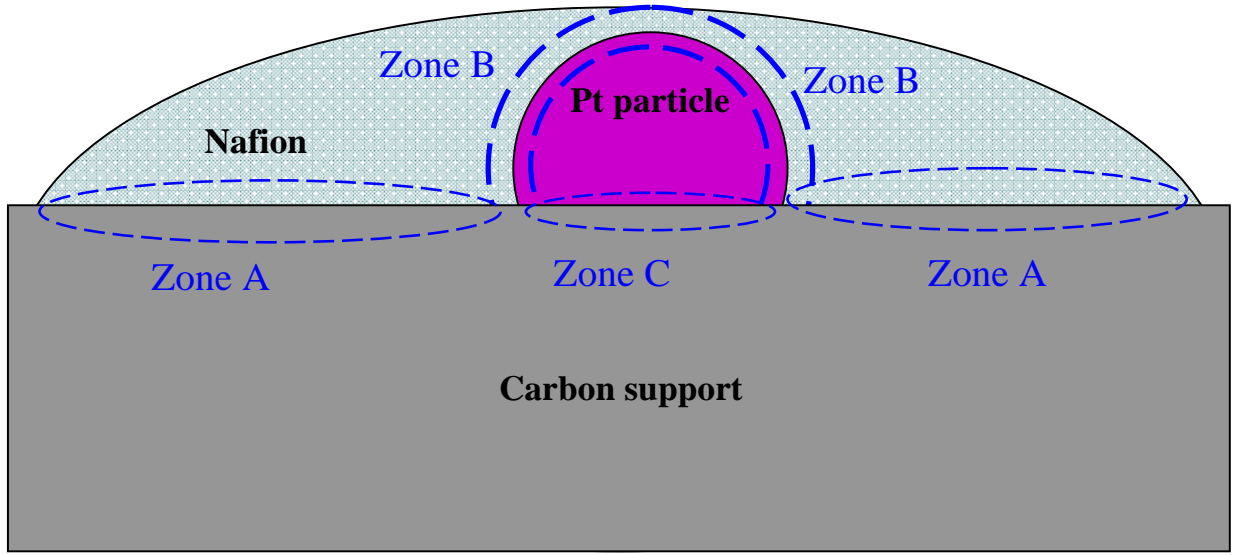

Figure 2: Schematic of the cathode catalyst layer

In the current carbon corrosion model the cathode catalyst layer is represented as three interfaces: zone A (Nafion/carbon interface), zone B (platinum/Nafion interface) and zone C (platinum/carbon interface). A schematic picture of the cathode catalyst layer consisting of these zones is shown in Figure 2. In each of these zones different sets of reactions are assumed to take place. These reactions are described below.

**Zone A (Interface between Nafion and carbon)**

We assume that the carbon surface consists of defect sites $C^*$ that are prone to reversible electrochemical oxidation forming an unstable $C$-$OH$ group at a potential of $E_{A1}$=0.2 V.

$$C^* + H_2O \leftrightarrow C-OH + H^+ + e^- \qquad E_{A1} = 0.2 \text{ V RHE} \qquad (1)$$

The unstable surface C-OH group can be further oxidized to form a stable C=O group at a potential of $E_{A2}$=0.8 V through the following reversible reaction.

$$C-OH \leftrightarrow C=O + H^+ + e^- \qquad E_{A2} = 0.8 \text{ V RHE} \qquad (2)$$

At a higher potential of $E_{A3} = 0.95\,\text{V}$, the unstable surface C-OH group is oxidized to form carbon dioxide that exits the system leaving behind a new defect site. This reaction is shown below.

$$\text{C} - \text{OH} + \text{H}_2\text{O} \rightarrow \text{C}^* + \text{CO}_2 + 3\text{H}^+ + 3\text{e}^- \qquad E_{A3} = 0.95\,\text{V RHE} \qquad (3)$$

Reaction 3 captures the carbon dioxide emission at higher voltages from the oxidation of the unstable surface groups which are not converted to the stable surface groups.

The spillover mechanism of carbon corrosion was also included in the model. This is represented by two other reactions involving ·OH radicals. OH groups are formed at Pt surface in the process of Pt electrochemical oxidation (discussed later in Zone B section). These OH groups can spill over to the carbon surface forming ·OH radical adsorbed at carbon surface.

$$\text{PtOH} + \text{C} \leftrightarrow \text{Pt} + \text{C}(\cdot\text{OH})_{\text{ads}} \qquad (4)$$

This absorbed ·OH radical can react with defect carbon site with formation of unstable C-OH groups. As the concentration of free defect sites is very low due to high activity of these sites we neglect the reaction of ·OH radicals with defect sites. Also ·OH radicals can react with the unstable carbon hydroxides forming carbon dioxide and creating a defect site as per the following reactions.

$$\text{C} - \text{OH} + \text{C}(\cdot\text{OH})_{\text{ads}} \leftrightarrow \text{CO}_2 + \text{C}^* + 2\text{H}^+ + 2\text{e}^- \qquad E_{A6} = 0.2V \qquad (5)$$

It should be noted that the ·OH radical attack on the surface oxide of carbon can result in carbon dioxide emission at low potentials, i.e. even at potentials just above 0.2 V. The reaction rate constants and the equilibrium potentials are not explicitly discussed in literature. Hence, most of the parameters used to represent these reactions are fitted to experimental data, to enable the prediction of the qualitative trends of the impact of potential change.

**Zone B (Interface between Nafion and platinum)**

On the surface between the platinum and Nafion, two sets of reactions take place: platinum oxidation and ·OH radical generation. The platinum oxidation occurs through the following two reactions with $Pt(OH)_{ads}$ acting as an intermediate.

$$Pt + H_2O \leftrightarrow Pt(OH)_{ads} + H^+ + e^- \quad E_{B1} = 0.7 \text{ V} \quad (6)$$

$$Pt(OH)_{ads} \leftrightarrow PtO_{ads} + H^+ + e^- \quad E_{B2} = 0.8 \text{ V} \quad (7)$$

The reaction rate parameters and the equilibrium potentials for reaction 6 and 7 are adopted from the literature[4,6]. The $Pt(OH)_{ads}$ intermediate can dissociate to generate ·OH radicals as per the following reaction.

$$Pt(OH)_{ads} \leftrightarrow Pt + C(\cdot OH)_{ads} \quad (8)$$

The ·OH radical that is generated on the platinum surface can spill over on to the carbon surface enabling carbon corrosion even at low potentials.

**Zone C (Interface between carbon and platinum)**

On the surface between the platinum and carbon, three sets of reactions occur. First, the platinum oxidation described by reactions (6) and (7) occur on the platinum surface in zone C. Second, the carbon oxidation reaction described by reaction (1), (2) and (3) occurs on the carbon surface in zone C. Third, the $Pt(OH)_{ads}$ reacts with the unstable surface oxide of carbon C-OH to form carbon dioxide and a carbon defect site as per the following reaction.

$$C-OH + Pt(OH)_{ads} \rightarrow C^* + CO_2 + Pt + 2H^+ + 2e^- \quad E_{C7} = 0.65 \text{ V} \quad (9)$$

This reaction allows for carbon dioxide emission during the step change in potential from low to a high value.

**Model equations**

The kinetics associated with each of the reactions (1) to (9) occurring in Zone A, Zone B and Zone C are captured in Table 1. State variables in these equations are listed in Table 2. The parameters in these equations are listed in the Appendix.

Table 1. The processes occurred in Zone A and reaction rates.

| Process | Reaction rate |
|---|---|
| Zone A | |
| C-OH hydroxide formation by reaction (1) | $R_{A1}(t) = k_{A1} \xi(\lambda) \left[ C(t) \cdot e^{\alpha_{A1}(V-E_{A1})/b_{A1}} - CX(t) \cdot e^{-(1-\alpha_{A1})(V-E_{A1})/b_{A1}} \right]$ |
| C=O oxides formation by reaction (2) | $R_{A2}(t) = k_{A2} \left[ CX(t) \cdot e^{\alpha_{A2}(V-E_{A2})/b_{A2}} - CY(t) \cdot \xi(\lambda) e^{-(1-\alpha_{A2})(V-E_{A2})/b_{A2}} \right]$ |
| Electrochemical $CO_2$ generation by reaction (3) | $R_{A3}(t) = k_{A3} CX(t) \cdot e^{\alpha_{A3}(V-E_{A3})/b_{A3}}$ |
| ·OH radicals adsorption on carbon by reaction (4) | $R_{A5}(t) = k_{A5}^{tot} \left[ c_{OH}^{bulk}(t) \dfrac{c_{OH}^{bulk,scal} L_{cath}}{\gamma_A} - k_{A5}^{eq} c_{OH}^{C,surf}(t) \right]$ |
| $CO_2$ generation by reaction (5) | $R_{A6}(t) = k_{A6} c_{OH}^{C,surf}(t) \cdot CX(t) e^{\alpha_{A6}(V-E_{A6})/b_{A6}}$ |
| Zone B | |
| PtOH formation by | $R_{B1}(t) = k_{B1} \left[ \xi(\lambda) \theta_{Pt}(t) e^{\alpha_{B1}(V-E_{B1}-r_{ox}\theta_{ox})/b_{B1}} - k_{rev} \theta_{PtOH}(t) e^{-(1-\alpha_{B1})(V-E_{B1})/b_{B1}} \right]$ |
| reaction (6) | |
| PtO formation by reaction (7) | $R_{B2}(t) = k_{B2}\left[\theta_{PtOH}(t)e^{\alpha_{B2}(V-E_{B2}-r_{ox}\theta_{ox})/b_{B2}} - k_{rev}\theta_{PtO}(t)e^{-(1-\alpha_{B2})(V-E_{B2})/b_{B2}}\right]$ |
| ·OH radicals desorption from Pt surface (8) | $R_{BX}(t) = k_{BX}^{tot}\left[\theta_{PtOH}(t) - k_{BX}^{eq}c_{OH}^{bulk}(t)\dfrac{c_{OH}^{bulk,scal}L_{cath}}{\gamma_B}\right]$ |
| Zone C | |
| PtOH formation by reaction (6) | $R_{C1}(t) = k_{B1}\left[\xi(\lambda)\theta_{Pt}(t)e^{\alpha_{B1}(V-E_{B1}-r_{ox}\theta_{ox})/b_{B1}} - k_{rev}\theta_{PtOH}(t)e^{-(1-\alpha_{B1})(V-E_{B1})/b_{B1}}\right]$ |
| PtO formation by reaction (7) | $R_{C2}(t) = k_{B2}\left[\theta_{PtOH}(t)e^{\alpha_{B2}(V-E_{B2}-r_{ox}\theta_{ox})/b_{B2}} - k_{rev}\theta_{PtO}(t)e^{-(1-\alpha_{B2})(V-E_{B2})/b_{B2}}\right]$ |
| C-OH hydroxide formation by reaction (1) | $R_{C4}(t) = k_{A1}\xi(\lambda)\left[C(t)\cdot e^{\alpha_{A1}(V-E_{A1})/b_{A1}} - CX(t)\cdot e^{-(1-\alpha_{A1})(V-E_{A1})/b_{A1}}\right]$ |
| C=O oxides formation by reaction (2) | $R_{C6}(t) = k_{A2}\left[CX(t)\cdot e^{\alpha_{A2}(V-E_{A2})/b_{A2}} - CY(t)\cdot \xi(\lambda)e^{-(1-\alpha_{A2})(V-E_{A2})/b_{A2}}\right]$ |
| $CO_2$ generation by reaction (3) | $R_{C5}(t) = k_{A3}CX(t)\cdot e^{\alpha_{A3}(V-E_{A3})/b_{A3}}$ |
| $CO_2$ generation by reaction (9) | $R_{C7}(t) = k_{C7}\theta_{PtOH}(t)CX(t)\cdot e^{\alpha_{c7}(V-E_{C7})/b_{C7}}$ |

Table 2. State variables

| State variables | |
|---|---|
| | |

| $C(t)$ | Fraction of carbon surface area free of oxides and OH radicals |
| --- | --- |
| $CX(t)$ | Fraction of carbon surface area covered with unstable hydroxide C-OH |
| $CY(t)$ | Fraction of carbon surface area covered with stable oxide C=O |
| $c_{OH}^{C,surf}(t)$ | Fraction of carbon surface area covered with OH radicals |
| $\theta_{Pt}(t)$ | Fraction of platinum surface area free of oxides |
| $\theta_{PtOH}(t)$ | Fraction of platinum surface area covered with PtOH |
| $\theta_{PtO}(t)$ | Fraction of platinum surface area covered with PtO |

A system of ordinary differential equations (ODEs) was developed using the kinetics equations presented Table 1.

In Zone A

$$\frac{d}{dt}\left(c_{OH}^{bulk}(t)\right) = \frac{1}{L_{cath}\varepsilon_N c_{OH}^{bulk,scal}}\left[\gamma_B R_{BX}(t) - \gamma_A R_{A5}(t)\right] \tag{10}$$

$$\frac{d}{dt}\left(c_{OH}^{C,surf}(t)\right) = R_{A5}(t) - \alpha_{C,defect} R_{A6}(t) \tag{11}$$

$$\frac{d}{dt}(C(t)) = -R_{A1}(t) - R_{A2}(t) + R_{A3}(t) + R_{A6}(t) \tag{12}$$

$$\frac{d}{dt}(CX(t)) = R_{A1}(t) - R_{A3}(t) - R_{A6}(t) \tag{13}$$

$$\frac{d}{dt}(CY(t)) = R_{A2}(t) \tag{14}$$

In Zone B

$$\frac{d}{dt}(\theta_M(t)) = -R_{B1}(t) + R_{BX}(t) \tag{15}$$

$$\frac{d}{dt}(\theta_{MOH}(t)) = R_{B1}(t) - R_{B2}(t) - R_{BX}(t) \tag{16}$$

$$\frac{d}{dt}(\theta_{MO}(t)) = R_{B2}(t) \tag{17}$$

In zone C

$$\frac{d}{dt}(C(t)) = -R_{C4}(t) - R_{C6}(t) + R_{C7}(t) + \frac{R_{C7}(t)}{\alpha_{Pt-CC}} \tag{18}$$

$$\frac{d}{dt}(CX(t)) = R_{C4}(t) - R_{C5}(t) - \frac{R_{C7}(t)}{\alpha_{Pt-CC}} \tag{19}$$

$$\frac{d}{dt}(CY(t)) = R_{C6}(t) \tag{20}$$

**Simulation results from the carbon corrosion model**

With the carbon corrosion model discussed above, we simulated the impact of various operating conditions on carbon dioxide emission as a measure of carbon corrosion. These results are presented below. Whenever experimental data is available we compare it against model predictions.

**Impact of potential hold**

At steady state, carbon dioxide emissions were measured in $H_2/N_2$ at anode and cathode of the fuel cell subscale system operating at a relative humidity of 95% and a temperature of 75°C. The experimental results along with model predictions for different potential holds are shown in

Figure 3. The $CO_2$ emission did not show a strong impact of potential at steady state. The observed variation was approximately ± 0.1 µg/min.

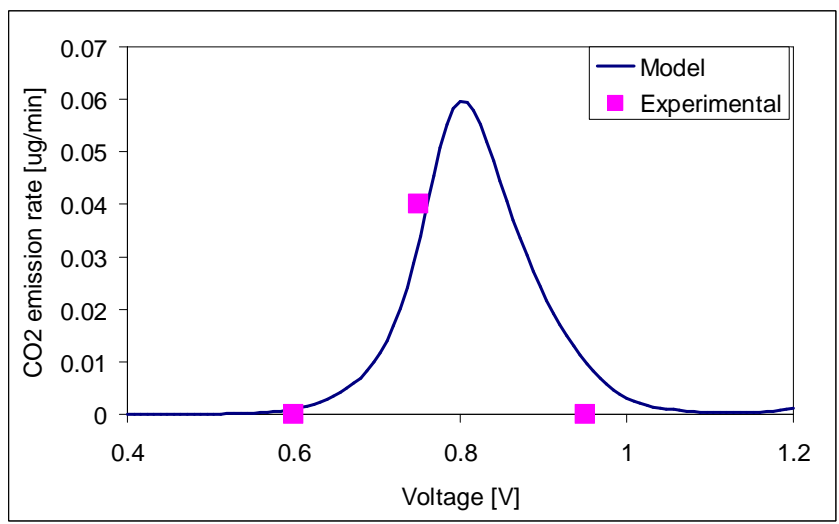

Figure 3: Steady state emission of $CO_2$ as a function of potential

The model prediction for carbon dioxide emission as function of potential had a peak in steady state emission at ~0.8 V. The quantitative emissions depend on the reaction rate constant for reaction (9) and the potential associated with the peak in $CO_2$ emissions is dependent on the equilibrium potential of reaction (2). It is important to note the emission rates predicted by the model are within the ± 0.1 µg/min of the $CO_2$ emission variation and these are interpreted to be fairly low $CO_2$ emissions from the fuel cell at steady state.

The steady state coverage fractions of various species on the platinum surface and carbon surface are shown in Figure 4 and Figure 5 respectively. It was seen that the decrease in C-OH at high potentials overlaps with the increase in $Pt(OH)_{ads}$ species which react to yield an increase in $CO_2$. The formation of the stable C=O at higher potentials subsequently decreases the $CO_2$ emissions at potentials greater than 0.8 V RHE. As seen from the figures, the interplay of the platinum surface species and carbon surface species is the source for carbon corrosion, as included in this model.

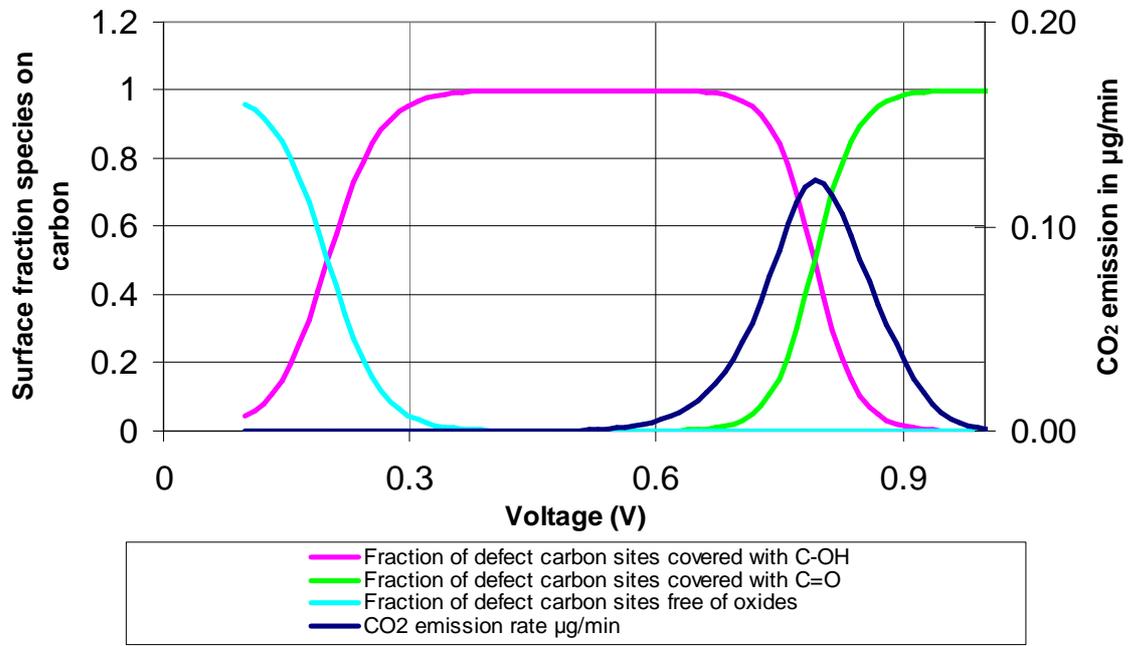

Figure 4: Steady state coverage of species on carbon surface as a function of potential

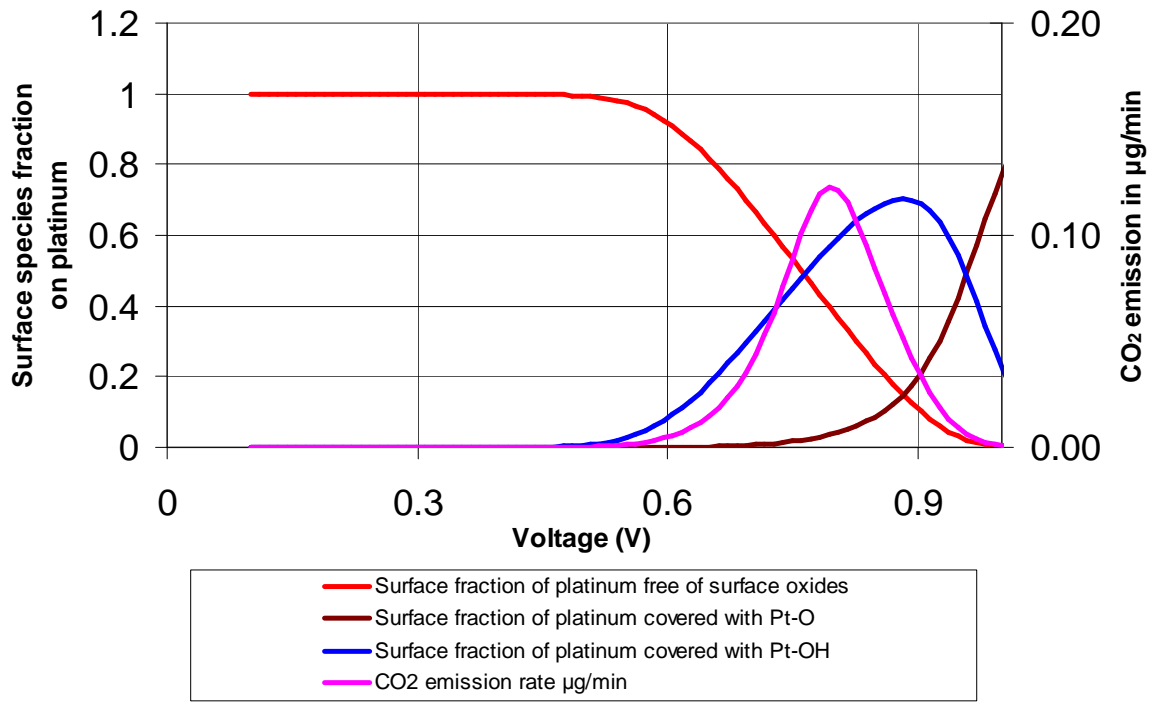

Figure 5: Steady state coverage of species on platinum surface as a function of potential

From the model presented in the previous section, there are three possible reactions (3, 5, and 9) that can produce carbon dioxide depending on the availability of reacting species and the potential. The steady state rates for these reactions are shown in Figure 6. The primary source of steady state carbon dioxide emission was reaction (9) that depends on availability of C-OH and $Pt(OH)_{ads}$.

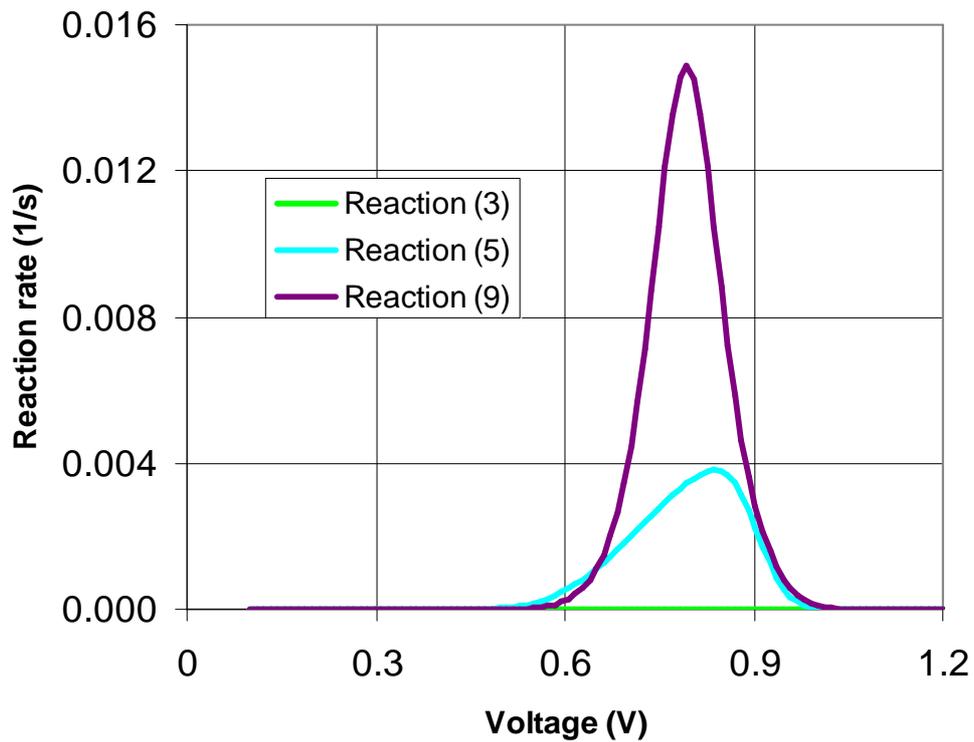

Figure 6: Steady state rates of reactions that produce $CO_2$

**Impact of step change in voltage**

Response of carbon corrosion to step changes in voltage were studied in a $H_2/N_2$ subscale system with the voltage changing from 0.1 V to 0.95 V and back to 0.1 V. The experimental measurement of carbon dioxide emission is shown in the left panel of Figure 7. We observed spikes in the carbon dioxide emission when there was a step change in voltage irrespective of the direction of change.

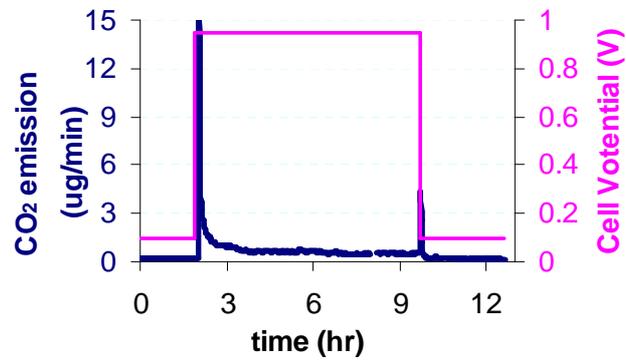

Figure 7: Experimental data of $CO_2$ emission response to step changes in voltage

Testing conditions: cell = 75ºC. Anode & cathode = 95%RH, 0.15SLPM $H_2$ & 0.36SPLM $N_2$ respectively

In Figure 8, the model prediction of $CO_2$ emission response to step up and step down in voltage is shown.

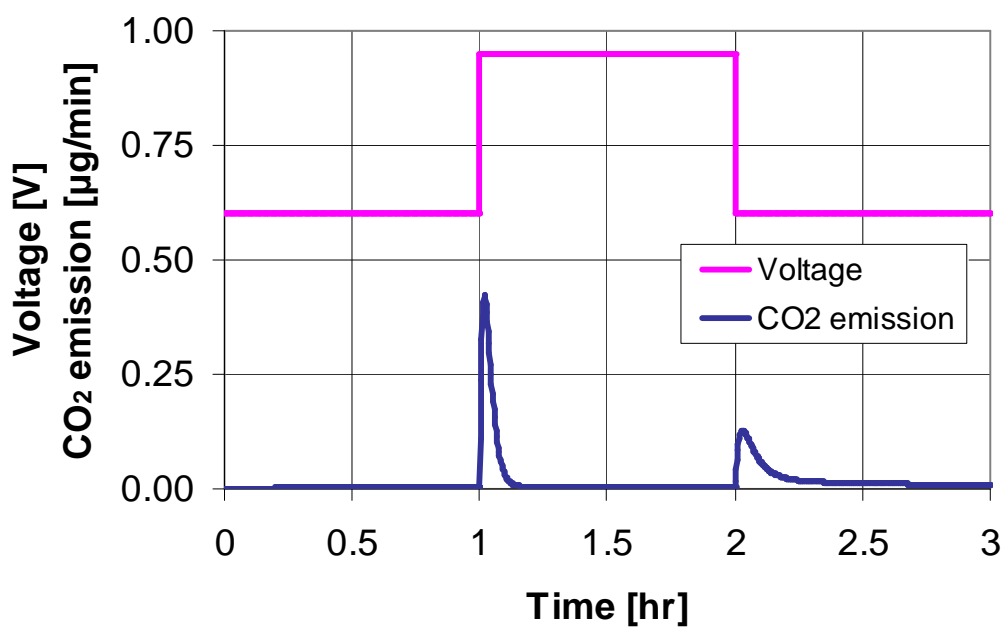

Figure 8: Model prediction of $CO_2$ emission response to step changes in voltage

The model prediction of the response of $CO_2$ emission was in qualitative agreement with the experimental observations. The model predicted a spike in carbon dioxide emission during a step up in the voltage because of the following sequences of events:

1. During the low potential hold C-OH forms and gets accumulated on the surface of the carbon.

2. As the voltage increases this accumulated C-OH is exposed to $Pt(OH)_{ads}$ and $(\cdot OH)_{ads}$ that form at high potentials.

3. By virtue of reactions (5) and (9), C-OH reacts with these species to release carbon dioxide until the surface fraction of C-OH settles down to a steady state value.

4. At voltages greater than 0.95 V, another contributor to carbon dioxide emission is direct oxidation of C-OH by water according to reaction (1).

The model predicts a spike in carbon corrosion during a step down in the voltage because of the following sequence of events:

1. During the high potential hold, the species Pt(OH)$_{ads}$ accumulates on the surface of platinum and the (·OH)$_{ads}$ accumulates on the surface of carbon.

2. As soon as the potential drops to a low value there is a sudden increase in the surface coverage of C-OH which reacts with the accumulated (·OH)$_{ads}$ and Pt(OH)$_{ads}$ to produce a spike in carbon dioxide emission.

3. As the accumulated (·OH)$_{ads}$ and Pt(OH)$_{ads}$ get consumed the carbon dioxide emission settles back to its steady state value.

At potential sweep down from 0.95 V to 0.6 V, $CO_2$ is produced by reaction (5) of C-OH with adsorbed ·OH radicals that were accumulated at carbon surface at high potential. The reaction (5) is anodic reaction and, therefore, the reaction rate constant decreases with decrease of potential according with Butler-Volmer equation. However, the reaction rate of reaction (5) is a product of the reaction rate constant, concentration of adsorbed ·OH radicals and concentration of C-OH at carbon surface. The concentration of C-OH increases with decrease of potential due to formation of C-OH from C=O by cathodic reaction (2). Thus, the total reaction rate of reaction (5) sharply increases for a short time, until the ·OH radicals are consumed. That results in transient increase of $CO_2$ emission.

Comparison of experimental data in Figure 7 and model predictions in Figure 8 shows similar qualitative trends between the $CO_2$ emissions measured experimentally and those predicted by the model. It is important to note that the magnitude of the spike predicted by the model is significantly lower than that observed in experiments.

The causes of discrepancy between the modeling and experimental results could include:

1. Under 0.95V potential hold, possible VOC originated from the gas source and/or from the stand (i.e. humidifier) could be oxidized into $CO_2$, leading to a higher $CO_2$ emission baseline in experimental curve.

2. The possible VOC or its derivatives may be absorbed on Pt and/or carbon surface during the constant potential hold (either low or high potentials), and be oxidized or reduced into $CO_2$ during the transient potential, which may significantly boost the magnitude of the spike in experimental curve.

**Impact of voltage cycling**

Next, we investigate the impact of voltage cycling over a period of 50 s consisting of 20 s hold at 0.6 V and 30 s hold at 0.95 V. Figure 9 shows the response of carbon dioxide emissions, in a $H_2/N_2$ subscale system, to voltage cycling between 0.6 V and 0.95 V.

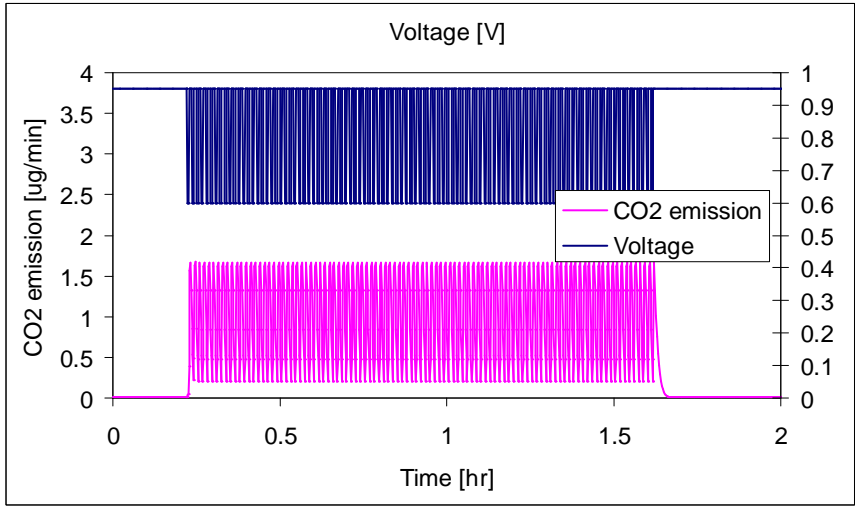

Figure 9: Model prediction of Carbon corrosion under 30s/20s voltage cycle between 0.6 V and 0.95 V

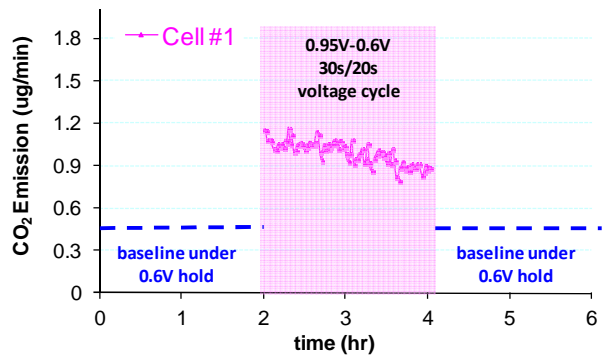

Figure 10. Experimental data of $CO_2$ emission response to 30 s / 20 s voltage cycle between 0.6 V and 0.95 V (75ºC, 95% RH, 0.15 slpm $H_2$ & 0.36 slpm $N_2$)

As seen from Figure 9, the carbon dioxide emission rate settled down during cycling to a steady state value greater than the emission rate corresponding to constant potential conditions, shown in Figure 3 . This was because cycling allowed for the co-existence of C-OH, Pt(OH)$_{ads}$ and (·OH)$_{ads}$ throughout entire cycling period. As a result, the average carbon dioxide emission during the cycling was about an order of magnitude greater than the carbon dioxide emission under potential hold conditions. This trend is in agreement with the experimental results shown in Figure 10. Consistent with the gap observed in the case of potential change, the magnitude of the spike due to change in potential from a steady state to the cyclic condition was not large.

**Impact of upper voltage during cycling**

Next, we studied the impact of upper voltage of the potential cycle on carbon corrosion. Figure 11 shows the comparison of the model predictions and the experimental measurements of the steady state carbon corrosion rate during cycling with lower potential fixed at 0.6 V and the upper potential varied as indicated. These simulations were carried out at an RH of 95%.

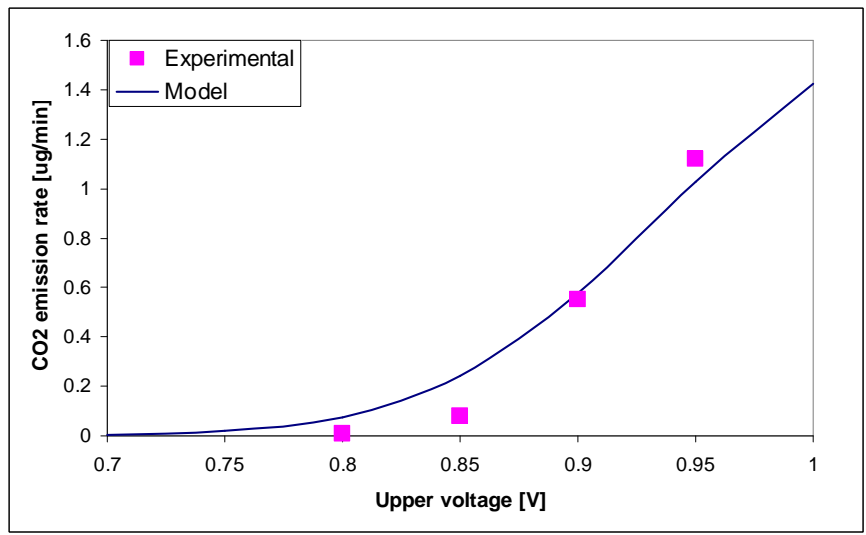

Figure 11: Comparison of the model predictions and experimental data on the impact of the upper voltage on carbon corrosion: 30 s / 20 s cycling, lower voltage = 0.6 V, at RH = 0.95

As seen in Figure 11, the steady state carbon dioxide emission rate during the cycling increased as the upper voltage increased from 0.7 V to 1.0 V. This is because the upper voltage increases to a value higher than equilibrium potential of Pt(OH)$_{ads}$ formation by reaction (6) which contribute to carbon dioxide emission via reaction (9). Reaction (3) also got activated at potentials greater than 0.9 V. Thus we observed a steady increase in the average carbon dioxide emission rate as the upper voltage in the voltage cycle went up. The model predictions were in qualitative agreement with the experimental measurements.

**Impact of lower voltage during cycling**

Next, we studied the impact of lower voltage of the voltage cycle on carbon corrosion. Figure 12 shows the comparison of the model predictions and the experimental measurements of the steady state carbon corrosion rate during cycling with upper potential fixed at 0.95 V and the lower potential varied as indicated. These simulations were carried out at an RH of 30%.

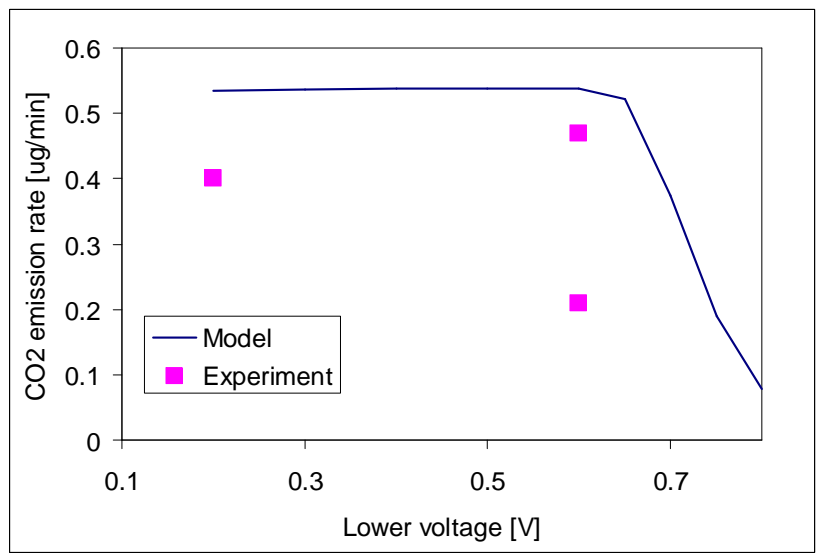

Figure 12: Comparison of the model predictions and experimental data on the impact of lower voltage on carbon corrosion: 30 s / 20 s cycling, upper voltage = 0.95 V at RH = 0.3

As shown in Figure 12, the steady state carbon dioxide emission rate during the cycling increased as the lower voltage decreased from 0.8 V to 0.6 V but the carbon dioxide emission rate increased marginally for voltages less than 0.6 V. This was because the propensity of C=O formation by consumption of C-OH via reaction (2) goes down as the voltage goes below 0.7 V leaving a higher surface coverage of C-OH. Since the carbon dioxide emission rate is dependent on the C-OH surface coverage, the average carbon dioxide emission rate increased as the lower voltage of the cycling decreased. The carbon dioxide emission rate plateaus at voltages less than 0.7 V because the carbon surface saturates with C-OH at low potentials. There is currently no mechanism built into the model to convert this C-OH to $CO_2$ at lower potentials.

**Impact of relative humidity during cycling**

Next, we studied the impact of relative humidity on carbon corrosion during voltage cycling. Figure 13 shows the comparison of the model predictions and the experimental measurements of the steady state carbon corrosion rate during voltage cycling between 0.95 V and 0.6 V with varied inlet gas relative humidity.

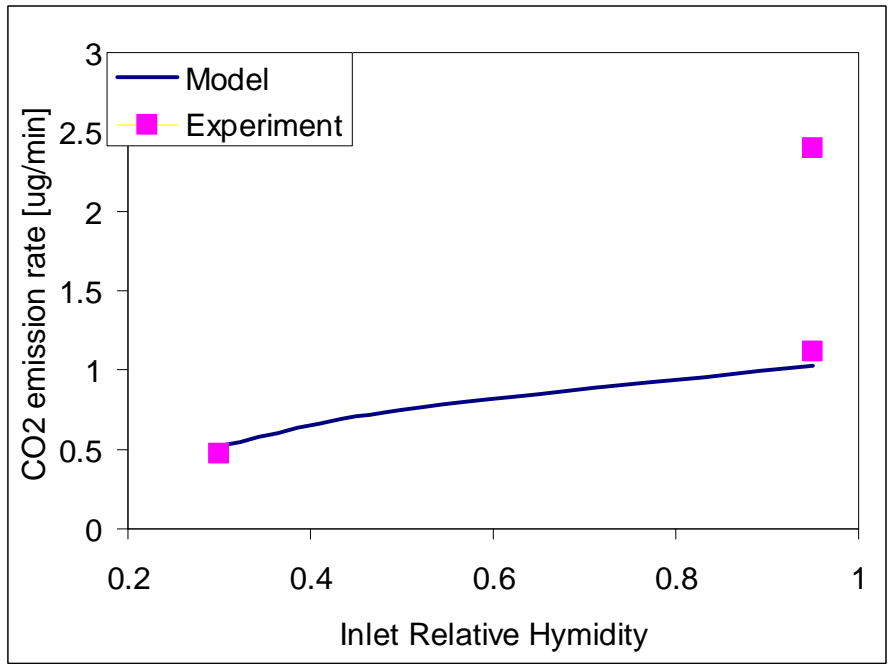

Figure 13: Impact of RH on carbon corrosion: 30s/20s cycling between 0.95 V and 0.6 V

The solid line in Figure 13 represents the model predictions and the squares are experimental data. The model predicts that the steady state carbon dioxide emission rate increased as the relative humidity increased up to 40% RH. This was because the formation of the unstable surface oxide of carbon C-OH depended on the availability of water. As the relative humidity went down, the propensity of the system to form C-OH went down and consequently the carbon corrosion rate also went down. Above 40% RH, the model predicts low impact of relative humidity on the carbon emission rate though the experimental data did show an increase in $CO_2$ emission. This is a gap in the existing model that needs further investigation.

**Conclusions**

A modeling framework that captured the effect of operational conditions on carbon corrosion was developed. The model predictions were in qualitative agreement with the carbon corrosion experiments. The interplay of the carbon surface oxides and radicals generated on platinum, the basic physics incorporated into the model, were key to predicting the carbon corrosion rates. Few key gaps were identified between the model predictions and experiments. Further model improvements are currently being made and will be reported in future.

**Appendix**

The nomenclature for the parameters in the model is presented below.

| Parameter | Value | Units; description |
|---|---|---|
| $E_{B1}$ | 0.7 V | Equilibrium potential of PtOH formation by reaction (6) |
| $n_{B1}$ | 1 | Electron transfer number of reaction (6) |
| $\alpha_{B1}$ | 0.5 | Transfer coefficient of reaction (6) |
| $k_{B1}$ | 1 s$^{-1}$ | Reaction rate of reaction (6) at equilibrium potential |
| $E_{B2}$ | 0.8 V | Equilibrium potential of PtO formation by reaction (7) |
| $n_{B2}$ | 1 | Electron transfer number of reaction (7) |
| $\alpha_{B2}$ | 0.5 | Transfer coefficient of reaction (7) |
| $k_{B2}$ | 1 s$^{-1}$ | Reaction rate of reaction (7) at equilibrium potential |
| $E_{A1}$ | 0.2 V | Equilibrium potential of COH formation by reaction (1) |

| $n_{A1}$ | 1 | Electron transfer number of reaction (1) |
|---|---|---|
| $\alpha_{A1}$ | 0.5 | Transfer coefficient of reaction (1) |
| $k_{A1}$ | $10^{-3}$ s$^{-1}$ | Reaction rate of reaction (1) at equilibrium potential |
| $E_{A2}$ | 0.6 | Equilibrium potential of CO formation by reaction (2) |
| $n_{A2}$ | 1 | Electron transfer number of reaction (2) |
| $\alpha_{A2}$ | 0.5 | Transfer coefficient of reaction (2) |
| $k_{A2}$ | $5*10^{-3}$ s$^{-1}$ | Reaction rate of reaction (2) at equilibrium potential |
| $E_{A3}$ | 0.9 | Equilibrium potential of $CO_2$ formation by reaction (3) |
| $n_{A3}$ | 3 | Electron transfer number of reaction (3) |
| $\alpha_{A3}$ | 0.5 | Transfer coefficient of reaction (3) |
| $k_{A3}$ | $3*10^{-6}$ s$^{-1}$ | Reaction rate of reaction (3) at equilibrium potential |
| $E_{C7}$ | 0.65 V | Equilibrium potential of $CO_2$ formation by reaction (9) |
| $n_{C7}$ | 2 | Electron transfer number of reaction (9) |
| $\alpha_{C7}$ | 0.5 | Transfer coefficient of reaction (9) |
| $k_{C7}$ | $10^{-1}$ s$^{-1}$ | Reaction rate of reaction (9) at equilibrium potential |
| $E_{A6}$ | 0.2 | Equilibrium potential of $CO_2$ formation by reaction (6) |
| $n_{A6}$ | 2 | Electron transfer number of reaction (6) |
| $\alpha_{A6}$ | 0.5 | Transfer coefficient of reaction (6) |
| $k_{A6}$ | $10^{-4}$ s$^{-1}$ | Reaction rate of reaction (6) at equilibrium potential |

| $k_{A5}^{tot}$ | $10^{-4}$ s$^{-1}$ | ·OH radicals absorption rate at carbon surface by reaction (4) |
|---|---|---|
| $k_{A5}^{eq}$ | 10 | Equilibrium constant for reaction (4) |
| $k_{BX}^{tot}$ | $10^{-3}$ s$^{-1}$ | ·OH radicals desorption rate from Pt surface by reaction (8) |
| $k_{BX}^{eq}$ | 0.1 | Equilibrium constant for reaction (8) |
| $k_{rev}$ | $3*10^{-1}$ | Reverse reaction deceleration in reactions (6) and (7) |
| $T_{ref}$ | 353 K | Reference temperature |
| $b_k$ | $b_k = \dfrac{RT_{ref}}{Fn_k}$ [V] | Tafel slope of k-th electrochemical reaction |
| $c_{OH,bulk}^{scal}$ | 1 mol/liter | Reference concentration |
| $\gamma_A$ | 0.01 | Concentration of "defect" (active) sites at carbon/Nafion interface |
| $\gamma_B$ | 1 | Concentration of active Pt sites at Pt/Nafion interface |
| $\alpha_{Pt-CC}$ | 1 | Number of Carbon atoms in contact with each Pt atom |
| $\alpha_{C,defect}$ | 0.005 | Fraction of "defect" carbon surface sites |
| $\varepsilon_N$ | 0.25 | Nafion volume fraction in electrode |
| $\xi$ | | Water activity |